\documentclass[aps,prl,superscriptaddress]{revtex4}
\usepackage{mathrsfs}
\usepackage{graphicx}
\usepackage{subfigure}
\usepackage{color}
\usepackage{epstopdf}

\usepackage[tbtags]{amsmath}
\usepackage{amssymb}
\usepackage{amsfonts}

\begin{document}

\title{Optimal GHZ Paradox for Three Qubits}

\author{Changliang Ren}
 \affiliation{Center of Quantum information, Chongqing Institute of Green and Intelligent Technology, Chinese Academy of Sciences, People's Republic of China}

\author{Hong-Yi Su\footnote{Correspondence to
H.Y.S. (hysu@mail.nankai.edu.cn).}}
 \affiliation{Theoretical Physics Division, Chern Institute of Mathematics, Nankai University,
 Tianjin 300071, People's Republic of China}

\author{Zhen-Peng Xu}
 \affiliation{Theoretical Physics Division, Chern Institute of Mathematics, Nankai University,
 Tianjin 300071, People's Republic of China}

\author{Chunfeng Wu}
\affiliation{Pillar of Engineering Product Development, Singapore University of Technology and Design, 8 Somapah Road, Singapore 487372}

\author{Jing-Ling Chen\footnote{Correspondence to
J.L.C. (chenjl@nankai.edu.cn).}}
 \affiliation{Theoretical Physics Division, Chern Institute of Mathematics, Nankai University,
 Tianjin 300071, People's Republic of China}
 \affiliation{Centre for Quantum Technologies, National University of Singapore,
 3 Science Drive 2, Singapore 117543}

\maketitle

\textbf{Quatum nonlocality as a valuable resource is of vital importance in quantum information processing. The characterization of the resource has been extensively investigated mainly for pure states, while relatively less is know for mixed states. Here we prove the existence of the optimal GHZ paradox by using a novel and simple method to extract an optimal state that can saturate the tradeoff relation between quantum nonlocality and the state purity. In this paradox, the logical inequality which is formulated by the GHZ-typed event probabilities can be violated maximally by the optimal state for any fixed amount of purity (or mixedness). Moreover, the optimal state can be described as a standard GHZ state suffering flipped color noise. The maximal amount of noise that the optimal state can resist is $50\%$. We suggest our result to be a step toward deeper understanding of the role played by the AVN proof of quantum nonlocality as a useful physical resource.}


\vspace{3mm}

The violation of Bell's inequality exhibits a conflict between local realism and quantum theory \cite{Bell,CHSH}. However, such a conflict has only been displayed \emph{statistically}. By devising effective logical arguments, an even sharper conflict between any local hidden variable model and quantum mechanical predictions can also be exhibited without resorting to inequalities. The most relevant examples of this line are Greenberger-Horne-Zeilinger (GHZ) \cite{GHZ 1989,GHZ 1990} and Hardy's proofs \cite{Hardy PRL 1993} of Bell nonlocality. These proofs are also referred to as the ``all-versus-nothing" (AVN) proof \cite{Mermin N Y Acad 755 (1995)}.

The AVN proof of, but not limited to, Bell nonlocality has attracted much attention and extensive results have been achieved both theoretically and experimentally. For instance, Cabello presented an AVN proof for two observers which holds for maximally entangled states \cite{Cabello1,Cabello2}; Scarani \textit{et.al} pointed out that any cluster state can display its nonlocality in the sense of GHZ paradox \cite{Scarani PRA 2005}; Cabello and Moreno presented the AVN proofs with $n$ qubits distributed between $m$ parties \cite{Cabello3}. On the other hand, the experimental tests of AVN proofs have been demonstrated by the two-photon hyperentanglement \cite{Cinelli,Yang,Barbieri} and by energy-time entanglement \cite{Vallone}. As shown in the literatures, the AVN proof not only opened ``a new chapter on the hidden variables problem" and made ``the strongest case against local realism since Bell's work", but also played an active role in quantum information science, such as quantum protocols to reduce communication complexity \cite{Cleve} and quantum key distribution protocols \cite{Liang}. Furthermore, the AVN proof has been shown to be effective as well in the studies of multipartite entanglement~\cite{Dur}, quantum steering \cite{Chen1,Chen2,Sun} and quantum contextuality \cite{Cabello4}.

However, most of the known results on the AVN proof are ideally based on pure states. In practical experiments, interaction between system and environment is unavoidable and hence pure entangled states inevitably become mixed states because of the effect of decoherence. So it is of significance to explore the AVN proof for mixed states. Although researchers' understanding of pure states has been meaningfully improved in recent years, mixed states has remained a notoriously difficult subject. Impressively, Ghirardi and Marinatto considered a nonlocality test without inequality in the case of mixed states \cite{Ghirardi PRA 73 032102 (2006)}. One year later, the AVN proof for multipartite mixed states have also been discussed \cite{Ghirardi PRA 74 022101 (2006),Ghirardi PRA 74 062107 (2006),Shi}.

In this work, we present the optimal three-qubit GHZ paradox in the sense that an optimal mixed state can be found such that the logical inequality, formulated by the GHZ-typed event probabilities in Ref.~\cite{Ghirardi PRA 73 032102 (2006)}, can be maximally violated for any fixed state mixedness.

\vspace{8pt}
\noindent{\bf Results}

\noindent \textbf{The optimal GHZ paradox.} To investigate the \emph{optimal} GHZ paradox of a three-qubit system, we start with the original case for the standard GHZ state
\begin{eqnarray}
\mid {\rm GHZ}\rangle=\frac{1}{\sqrt{2}}(\mid000\rangle+\mid111\rangle),\label{GHZ}
\end{eqnarray}
where $\mid0\rangle$ and $\mid1\rangle$ are the eigenstates of the
Pauli matrix $\sigma_z$ associated to the eigenvalues $+1$ and $-1$,
respectively. Consider a set of four mutually commutative observables $\sigma_{1x}\otimes\sigma_{2x}\otimes\sigma_{3x}$, $\sigma_{1y}\otimes\sigma_{2x}\otimes\sigma_{3y}$, $\sigma_{1y}\otimes\sigma_{2y}\otimes\sigma_{3x}$, and $\sigma_{1x}\otimes\sigma_{2y}\otimes\sigma_{3y}$, where $\sigma_{1x}$\ is defined as the Pauli matrix $\sigma_{x}$ measured on the $1$-st qubit (similarly for the others), and state (\ref{GHZ}) is the common eigenstate of these four
operators, with the eigenvalues being $+1$, $-1$, $-1$, $-1$, respectively.

However, as shown in Ref.~\cite{GHZ 1990}, a contradiction arises if one tries to interpret the quantum result with local hidden variable (LHV) models, in which each local observable has two definite values, $+1$ and $-1$, even before the measurements. Specifically, we denote the supposedly definite values of $\sigma_{1x}$, $\sigma_{2y}$, $\cdots$\ as $v_{1x}$,
$v_{2y}$, $\cdots$, then a product of the last three observables, according to LHV models, yields $v_{1y}^{2}v_{2y}^{2}v_{3y}^{2}v_{1x}v_{2x}v_{3x}=-1$, in sharp contradiction to the first observable $v_{1x}v_{2x}v_{3x}=+1$.

From the point of view of experiments, environment-induced noise is generally unavoidable, and the efficiency of generating three-qubit entangled states in the laboratory is usually below $90\%$. Hence it is important and highly nontrivial to take account of the case of mixed states when studying the GHZ paradox. Ghirardi and Marinatto \cite{Ghirardi PRA 74 022101 (2006)} demonstrated that the GHZ proof of nonlocality exists for a mixed state $\rho$ if the following inequality
\begin{equation}\label{INEQ}
3-q_1-q_2-q_3-q_4\geq 0
\end{equation}
is violated, where $q_i$ is defined as the event probability for each observable mentioned above that happens with certainty:
\begin{eqnarray}
q_1 =P_{\rho}(\sigma_{1x}\otimes\sigma_{2x}\otimes\sigma_{3x}=+1),\label{GHZ paradox1}\\
q_2 =P_{\rho}(\sigma_{1y}\otimes\sigma_{2x}\otimes\sigma_{3y}=-1),\label{GHZ paradox2}\\
q_3 =P_{\rho}(\sigma_{1y}\otimes\sigma_{2y}\otimes\sigma_{3x}=-1),\label{GHZ paradox3}\\
q_4 =P_{\rho}(\sigma_{1x}\otimes\sigma_{2y}\otimes\sigma_{3y}=-1)\label{GHZ paradox4}.
\end{eqnarray}
The degree of violation of (\ref{INEQ}) exhibits the degree of nonlocality.

Thus one may ask: What is the optimal state that violates (\ref{INEQ}) maximally so as to show the largest degree of nonlocality? To address the problem we use the notion of linear entropy \cite{Wei}, which is a measure of state mixedness and computed as
\begin{equation}\label{entropy}
\varepsilon(\rho)=\frac{d}{d-1}(1-\mathrm{Tr}(\rho^2)),
\end{equation}
where the fraction is for normalization, with $d=2^N$ and $N$ being the number of qubits. Put more precisely, the problem now becomes: Which state, for a fixed value of its linear entropy, can achieve the maximal violation of (\ref{INEQ})? We usually need to solve a optimization problem by evaluating all states in the Hilbert space. The computation complexity increases very rapidly as the dimension of Hilbert space increases. Here we focus on three qubits and present a simple and rigorous method to obtain the optimal state, which reads
\begin{equation}\label{opt density3}
\rho_{\mathrm{DM}}^{\mathrm{opt}}=\left(  \begin{array}{cccccccc}
f_1&0 &0 &0 &0&0 &0&f_1\\
0&f &0 &0 &0&0 &0&0\\
0&0 &f &0 &0&0 &0&0\\
0&0 &0 &f &0&0 &0&0\\
0&0 &0 &0 &f&0 &0&0\\
0&0 &0 &0 &0&f &0&0\\
0&0 &0 &0 &0&0 &f&0\\
f_1&0 &0 &0 &0&0 &0&f_1
\end{array}\right),
\end{equation}
where $f=\frac{(1-2f_1)}{6}$, and subscript $DM$ indicates that it is the form of physical density matrix giving maximal entropy for a given amount of violation.

The rest of the paper will be contributed to a proof of (\ref{opt density3}). But before the proof, we would like to give some discussions on the physical aspect of the state. Let us rewrite the optimal state (\ref{opt density3}) as
\begin{align}\label{opt density4}
&\rho_{\mathrm{DM}}^{\mathrm{opt}}=2f_1\mid GHZ\rangle\langle GHZ\mid\nonumber\\&+f\mid 100\rangle\langle 100\mid+f\mid 010\rangle\langle 010\mid+f\mid 001\rangle\langle 001\mid\nonumber\\&+f\mid 110\rangle\langle 110\mid+f\mid 011\rangle\langle 011\mid+f\mid 101\rangle\langle 101\mid.
\end{align}
Its linear entropy, according to Eq.(\ref{entropy}), equals
\begin{equation}\label{opt entropy}
\varepsilon(\rho_{\mathrm{DM}}^{\mathrm{opt}})=\frac{8}{7}(1-4f_1^2-6f^2).
\end{equation}
It is clear that the optimal state (\ref{opt density4}) is a mixture of the standard GHZ state and flipped noise equally upon six of eight bases.

Given the condition Eq.(\ref{definition2}) (see below), we must have $|f_1|\leq\frac{1}{2}$. Hence our GHZ proof is valid only when $\frac{1}{4}<f_1\leq\frac{1}{2}$. When $f_1=\frac{1}{2}$, the optimal state is reduced to the standard GHZ state; when $f_1<\frac{1}{4}$, which corresponds to the case where the amount of flip noise weighs over $50\%$, the state could be compatible with a LHV model. Thus $50\%$ is the upper bound of the amount of flipped noise that can be resisted by the optimal state.

\vspace{5mm}
\noindent \textbf{Proof of the optimal state.} A three-qubit state can be represented as \cite{Horodecki1,Horodecki2}
\begin{align}\label{state2}
\rho=& \frac{1}{8} \sum_{r,s,t=x,y,z,0} p_{rst} \sigma_r\otimes \sigma_s
\otimes \sigma_t,
\end{align}
where $p_{rst}$'s are real coefficients, $\sigma_0$ denotes the identity matrix, and $p_{000}=1$ for consistency. Of course, $p_{rst}$ should be subject to some constraints: (i) $\rho$ should be positive semi-definite, and (ii) the trace of $\rho$ should be unity.

The linear entropy of state (\ref{state2}) equals the sum of square of each coefficient, i.e.,
\begin{align}\label{entropy2}
\varepsilon(\rho)=\frac{8}{7}(1-\sum_{i,j=1}^{8}\rho_{ij}^2)=1-\frac{1}{7}\sum_{r,s,t} p_{rst} ^2.
\end{align}
It is evident that the state that gives the maximal violation of (\ref{INEQ}) for a fixed linear entropy is equivalent to the state that gives the maximal linear entropy for a fixed quantum violation.

For the given
state of the form (\ref{state2}), the four event probabilities defined by Eq.(\ref{GHZ paradox1}-\ref{GHZ paradox4}) can be obtained as
\begin{eqnarray}\label{GHZ paradox2}
q_1=\frac{1+p_{111}}{2},\label{1}
\\
q_2=\frac{1-p_{212}}{2},\label{2}
\\
q_3=\frac{1-p_{221}}{2},\label{3}
\\
q_4=\frac{1-p_{122}}{2},\label{4}
\end{eqnarray}
which imply that
\begin{equation}\label{definition2}
|p_{111}|\leq 1, |p_{212}|\leq 1, |p_{221}|\leq 1, |p_{122}|\leq 1,
\end{equation}
and substituting (\ref{1}-\ref{4}) into (\ref{INEQ}), we obtain
\begin{equation}\label{INEQ1}
1-\frac{(p_{111}-p_{212}-p_{221}-p_{122})}{2}\geq 0,
\end{equation}
which implies that the violation only depends on four
coefficients $p_{111}$, $p_{212}$, $p_{221}$, $p_{122}$. Only when $p_{111}-p_{212}-p_{221}-p_{122}>2$ can inequality (\ref{INEQ1}) be violated. Note that the algebraic maximum of $p_{111}-p_{212}-p_{221}-p_{122}$ is $4$. Intriguingly, the real part of matrix elements $\rho_{18}$ and $\rho_{81}$ can be expressed by $\frac{p_{111}-p_{212}-p_{221}-p_{122}}{8}$. For simplicity, we hereafter denote $p_{111}-p_{212}-p_{221}-p_{122}$ by $8f_1$, and then inequality (\ref{INEQ1}) becomes
\begin{eqnarray}
1-4f_1\geq 0.
\end{eqnarray}

According to Eq.(\ref{entropy2}), for a fixed value of $f_1$, matrix elements that do not contribute to $f_1$ should be set to zero in order to maximize the linear entropy, i.e., the less the irrelevant matrix elements there are, the higher the linear entropy becomes. Therefore the following form of matrix seems to be the best solution: (see \textbf{Methods})
\begin{equation}\label{opt density1}
\rho_{\mathrm{M}}^{\mathrm{opt}}=\left(  \begin{array}{cccccccc}
0&0 &0 &0 &0&0 &0&f_1\\
0&0 &0 &0 &0&0 &0&0\\
0&0 &0 &0 &0&0 &0&0\\
0&0 &0 &0 &0&0 &0&0\\
0&0 &0 &0 &0&0 &0&0\\
0&0 &0 &0 &0&0 &0&0\\
0&0 &0 &0 &0&0 &0&0\\
f_1&0 &0 &0 &0&0 &0&0
\end{array}\right),
\end{equation}
except the fact that it is not a density matrix. 

As we know, a real density matrix is positive semi-definite and its trace is unity. According to the sufficient and necessary condition of positive semi-definite matrix and similar discussion above, it is easily and rigorously to obtain the form of optimal positive semi-definite matrix that has the maximal entropy for a given amount of violation $f_1$, which can be expressed as
\begin{equation}\label{opt density2}
\rho_{\mathrm{PM}}^{\mathrm{opt}}=\left(  \begin{array}{cccccccc}
f_1&0 &0 &0 &0&0 &0&f_1\\
0&0 &0 &0 &0&0 &0&0\\
0&0 &0 &0 &0&0 &0&0\\
0&0 &0 &0 &0&0 &0&0\\
0&0 &0 &0 &0&0 &0&0\\
0&0 &0 &0 &0&0 &0&0\\
0&0 &0 &0 &0&0 &0&0\\
f_1&0 &0 &0 &0&0 &0&f_1
\end{array}\right)
\end{equation}
where one has to set $\rho_{11}=\rho_{88}=f_1$. Compared with positive semi-definite matrix, density matrix has one more restrict, that is $\sum^8_{i=1}\rho_{ii}=1$. Obviously, only when $f_1=1/2$, Eq.(\ref{opt density2}) represents a real physical density matrix, which means the optimal density matrix can reach that of positive semi-definite matrix Eq.(\ref{opt density2}).
In this case it is nothing but the maximally entangled pure state which given the maximal violation of Eq.(\ref{INEQ1}). However, when $f_1<1/2$, the optimal positive semi-definite matrix is not a density matrix. To obtain the form of the optimal density matrix, the diagonal matrix elements should have nonzero terms and the sum of them must be 1, hence, in computing the linear entropy we note that
\begin{align}
\mathrm{Tr}(\rho^2)&= 2f_1^2+\sum^8_{i=1}\mid\rho_{ii}\mid^2 \nonumber\\&\geq 4f_1^2+\sum^7_{i=2}\mid\rho_{ii}\mid^2\nonumber\\&\geq 4f_1^2+\frac{(1-2f_1)^2}{6}. \nonumber
\end{align}
That is, due to the geometric-mean inequality, the maximal linear entropy can be obtained when $\rho_{11}=\rho_{88}=f_1$ and $\rho_{22}, \rho_{33}, \cdots,\rho_{77}$ are equal to each other, as is precisely the case of (\ref{opt density3}). Hence state (\ref{opt density3}) is indeed the optimal state that has the largest degree of nonlocality.

\vspace{8pt}
\noindent{\bf Discussion}

In conclusion, we have proved the optimal GHZ paradox by finding the hull of quantum states that saturate the trade-off relation between the linear entropy and the quantum violation of (\ref{INEQ}). The optimal state can be considered as the standard GHZ state suffering flipped color noise, and we have shown that the exhibition of the GHZ paradox for the optimal state depends on the amount of the noise: the stronger the noise is, the less nonlocality the optimal state has. When the amount of noise is over $50\%$, the optimal state does not bear the  nonlocality that violates inequality (\ref{INEQ}). The method we use in the present paper provides a particularly new perspective to understand the GHZ paradox for mixed states, and our results may have potential applications in quantum information processing.

%

\vspace{8pt}
\noindent{\bf Methods}

\noindent\textbf{ Derivation of $\rho_{\mathrm{M}}^{\mathrm{opt}}$.}
As we know, all matrix elements $\rho_{ij}$ in Eq.(\ref{state2}) are the combinations of $p_{rst}$'s. Because $\rho_{18}$ and $\rho_{81}$ depend on the four coefficients $p_{111}$, $p_{212}$, $p_{221}$, $p_{122}$, we only need to consider the matrix elements containing coefficients $p_{111}$, $p_{212}$, $p_{221}$, $p_{122}$, and set the others zero. Obviously, except the anti-diagonal matrix elements, the remaining matrix elements do not depend on such coefficients, so they can be set to zero.

Then we analyze whether the anti-diagonal matrix elements, except $\rho_{18}$ and $\rho_{81}$, can be zero. We solve the following equations
\begin{eqnarray}
\rho_{18}=p_{111}-p_{212}-p_{221}-p_{122}=f_1,\\
\rho_{27}=p_{111}-p_{212}+p_{221}+p_{122}=0,\\
\rho_{36}=p_{111}+p_{212}-p_{221}+p_{122}=0,\\
\rho_{45}=p_{111}+p_{212}+p_{221}-p_{122}=0.\\
\end{eqnarray}
The solutions are found to be $p_{111}=\frac{f_1}{4}$ and $p_{212}=p_{221}=p_{122}=-\frac{f_1}{4}$. Hence the matrix (\ref{opt density1}) is indeed the form of matrix that has the maximal linear entropy for a fixed amount of violation $f_1$.

\vspace{8pt}

{\bf Acknowledgements}

J.L.C. is supported by National Basic Research Program (973 Program) of China under Grant No.\ 2012CB921900 and the NSF of China (Grant Nos.\ 10975075 and 11175089). C.R. thanks Chern Institute of Mathematics for invited visiting and acknowledges supported by Youth Innovation Promotion Association (CAS) No.2015317.

\vspace{2pt}

{\bf Author contributions}

J.L.C. initiated the idea. C.R., H.Y.S., Z.P.X., and J.L.C. provided the proof. C.R., H.Y.S., and C.W. wrote the main manuscript text. All authors reviewed the manuscript.

\vspace{2pt}

{\bf Additional information}

Competing financial interests: The authors declare no competing financial interests.

\end{document}